\documentclass[preprint,12pt]{aastex}

\shortauthors{Winn et al.~2007}
\shorttitle{System Parameters of HAT-P-1}

\begin{document}

%
\def\ltsima{$\; \buildrel < \over \sim \;$}
\def\lsim{\lower.5ex\hbox{\ltsima}}
\def\gtsima{$\; \buildrel > \over \sim \;$}
\def\gsim{\lower.5ex\hbox{\gtsima}}
\def\lam{\lambda=-1\fdg4 \pm 1\fdg1}
%

\bibliographystyle{apj}

\title{
The Transit Light Curve Project.\\
VII.~The Not-So-Bloated Exoplanet HAT-P-1b
}

\author{
Joshua N.\ Winn\altaffilmark{1},
Matthew J.\ Holman\altaffilmark{2},
Gaspar \'{A}.\ Bakos\altaffilmark{2,3},
Andr\'{a}s P\'{a}l\altaffilmark{4},\\
John Asher Johnson\altaffilmark{5},
Peter K.\ G.\ Williams\altaffilmark{5},
Avi Shporer\altaffilmark{6},
Tsevi Mazeh\altaffilmark{6},\\
Jos\'{e} Fernandez\altaffilmark{2,7},
David W.\ Latham\altaffilmark{2},
Micha\"{e}l Gillon\altaffilmark{8}
}

\altaffiltext{1}{Department of Physics, and Kavli Institute for
  Astrophysics and Space Research, Massachusetts Institute of
  Technology, Cambridge, MA 02139, USA}

\altaffiltext{2}{Harvard-Smithsonian Center for Astrophysics, 60
  Garden Street, Cambridge, MA 02138, USA}

\altaffiltext{3}{Hubble Fellow}

\altaffiltext{4}{Department of Astronomy, E\"{o}tv\"{o}s Lor\'{a}nd
  University, Pf.~32, H-1518 Budapest, Hungary}

\altaffiltext{5}{Department of Astronomy, University of California,
  Mail Code 3411, Berkeley, CA 94720, USA}

\altaffiltext{6}{Wise Observatory, Raymond and Beverly Sackler Faculty
  of Exact Sciences, Tel Aviv University, Tel Aviv 69978, Israel}

\altaffiltext{7}{Department of Astronomy, Pontificia Universidad
  Cat\'{o}lica, Casilla 306, Santiago 22, Chile}

\altaffiltext{8}{Observatoire de l'Universit\'e de Gen\`eve, 1290
  Sauverny, Switzerland; and Institut d'Astrophysique et de
  G\'eophysique, Universit\'e de Li\`ege, 4000 Li\`ege, Belgium}

\begin{abstract}

  We present photometry of the G0 star HAT-P-1 during six transits of
  its close-in giant planet, and we refine the estimates of the system
  parameters. Relative to Jupiter's properties, HAT-P-1b is $1.20\pm
  0.05$~times larger and its surface gravity is $2.7\pm 0.2$~times
  weaker. Although it remains the case that HAT-P-1b is among the
  least dense of the known sample of transiting exoplanets, its
  properties are in accord with previously published models of
  strongly irradiated, coreless, solar-composition giant planets. The
  times of the transits have a typical accuracy of 1~min and do not
  depart significantly from a constant period.

\end{abstract}


\section{Introduction}

More than 12 years have passed since the surprising discovery of ``hot
Jupiters'': giant planets around Sun-like stars with orbital periods
smaller than $\sim$10~days~(Mayor \& Queloz 1995; Butler et
al.~1997). These objects, of which about 50 are known (see, e.g.,
Butler et al.~2006), probably formed at larger orbital distances and
migrated inwards through processes that are not yet fully understood.
Hot Jupiters are also interesting because they are more likely to
transit their parent stars than more distant planets. Transits are
highly prized because they permit the determination of the planetary
radius and mass (Henry et al.~2000, Charbonneau et al.~2000), the
infrared planetary spectrum (Richardson et al.~2007, Grillmair et
al.~2007) and longitudinal brightness profile (Knutson et al.~2007a),
the stellar obliquity (Winn et al.~2007a), and much more. This helps
to explain why so many groups around the world are conducting
wide-field photometric surveys for planetary transits. Over a dozen
cases of transiting exoplanets have been identified in this manner,
with the rest having been found as a by-product of Doppler planet
surveys (see Charbonneau et al.~2006 for a recent review).

Recently, the Hungarian-made Automated Telescope Network (HATNet)
announced the discovery of HAT-P-1b, a giant planet that orbits one
member of a G0/G0 stellar binary (Bakos et al.~2007). This planet was
notable for being among the largest and least dense of all the planets
for which such measurements have been made---both inside and outside
the Solar system---and therefore an interesting test case for models
of planetary atmospheres and interiors.

However, while the data presented by Bakos et al.~(2007) was certainly
good enough to clinch the case for planethood and to provide useful
estimates of the system parameters, it is possible and desirable to
improve the accuracy of those parameters with repeated,
high-precision, ground-based transit photometry. This is one goal of
our Transit Light Curve (TLC) project, which has been described at
greater length elsewhere (see, e.g., Holman et al.~2006, Winn et
al.~2007b).

This paper presents our results for the HAT-P-1 system. The next
section describes the observations.  In \S~3, we describe the
parameteric model that was fitted to the data, and in \S~4 we present
the results for the planetary, stellar, and orbital parameters, as
well as the transit times. At the end of this paper we discuss the
significance of the refined radius measurement.

\section{Observations and Data Reduction}

Our observations took place in late 2006, using telescopes at three
different observatories. We observed 6 distinct transits and produced
7 independent light curves.

We observed the transits of UT~2006~Sep~18, Sep~27, and Oct~6 with the
1.2m telescope at the Fred L.\ Whipple Observatory on Mt.\ Hopkins,
Arizona. We used the 4096$^2$ KeplerCam CCD, which has a $23\farcm1
\times 23\farcm 1$ field of view. We employed $2\times 2$ binning,
giving a scale of $0\farcs 68$ per binned pixel, a readout and setup
time of 11~s, and a typical readout noise of 7 e$^-$ per binned
pixel. We used a Sloan $z$ filter in order to minimize the effect of
atmospheric extinction on the relative photometry, and the effects of
stellar limb darkening on the transit light curve. We kept the image
registration as constant as possible. We also obtained dome-flat and
bias exposures at the beginning and the end of each night. On Sep~18,
the sky conditions were photometric and the seeing varied from
$1\farcs7$ to $2\farcs1$. We used an exposure time of 15~s. The night
of Sep~27 began with patchy clouds and large transparency variations,
but the rest of the night was clear. The seeing varied between
$1\farcs7$ and $2\farcs4$, and we again used an exposure time of
15~s. Most of Oct~6 was lost to clouds, although we did manage to
observe the egress in $1\farcs8$ seeing, using an exposure time of
10~s.

We observed the transits of UT~2006~Sep~1, UT~2006~Sep~10 and
UT~2006~Sep~18 using the Nickel~1m telescope at Lick Observatory on
Mt.\ Hamilton, California. The only night when a complete transit
could be observed was Sep~18. We used the Dewar \#2 direct imaging
detector, which is a 2048$^2$ Lawrence Labs CCD with a $6\farcm 1
\times 6\farcm 1$ field of view. For our observations we used $2\times
2$ binning (0\farcs 36 per binned pixel), and read out only a $1450
\times 500$ pixel subregion of the chip to decrease the readout
time. Setup and readout time took about 10 seconds per exposure, with
a typical read noise of 11.8 e$^-$ per binned pixel. We observed
through a ``Gunn $Z$'' filter (Pinfield et al.~1997). To draw out the
exposure time and to spread the light from stars over more pixels, we
defocused the telescope until the stellar images had a full-width at
half-maximum (FWHM) of about 6 pixels. The exposure time ranged from
10 to 40 seconds, depending on seeing and transparency. All nights
were fairly clear with $1\farcs 0$--$1\farcs 5$ seeing. On Sep~18,
near the transit midpoint, the star passed within a few degrees of the
zenith and autoguiding failed. The data from that time period were
excised.

We observed the transits of UT~2006~Sep~14, UT~2006~Nov~20, and
UT~2006~Nov~29 using the 1m telescope at Wise Observatory, in Israel.
We used a Tektronix 1024$^2$ back-illuminated CCD detector, giving a
pixel scale of $0\farcs 7$ and a field of view of $11\farcm 9 \times
11\farcm 9$. We observed through a Johnson $I$ filter, the reddest
optical band available on the camera. On Sep~14 and Nov~20, weather
conditions were poor, with patchy clouds. Because the data from those
nights were of much lower quality than the other data presented in
this paper, in what follows we describe only the data from
2006~Nov~29.  The night was not photometric, and the measured stellar
fluxes varied by about 20\% over the course of the night. The exposure
time was 15~s, and the FWHM of stellar images was about $1\farcs8$
(2.5~pixels).

We used standard IRAF\footnote{ The Image Reduction and Analysis
  Facility (IRAF) is distributed by the National Optical Astronomy
  Observatories, which are operated by the Association of Universities
  for Research in Astronomy, Inc., under cooperative agreement with
  the National Science Foundation.  } procedures for the overscan
correction, trimming, bias subtraction, and flat-field division. We
performed aperture photometry of HAT-P-1 and 4-8 nearby stars,
depending on the telescope. The sum of the fluxes of the comparison
stars was taken to be the comparison signal. The light curve of
HAT-P-1 was divided by the comparison signal, and then divided by a
constant to give a unit mean flux outside of transit.

We then assessed residual systematic effects by examining the
correlation between the out-of-transit flux and some external
variables: time, airmass, the shape parameters of stellar images, and
the pixel position of HAT-P-1. For the FLWO data, the flux variations
were most strongly correlated with airmass; for the Lick data, the
strongest correlations were with the pixel coordinates, especially the
row number; and for the Wise data, there were correlations with both
airmass and the FWHM of stellar images (which were themselves strongly
correlated). We solved for the zero point and slope of the strongest
correlation as part of the fitting process described in the next
section.

Figures~1 and 2 show the final light curves. The bottom panel of
Fig.~2 is a phase-averaged composite of the 3 best light
curves. Table~1 provides the final photometry, after correction of the
residual systematic effects.

\begin{figure}[p]
\epsscale{0.85}
\plotone{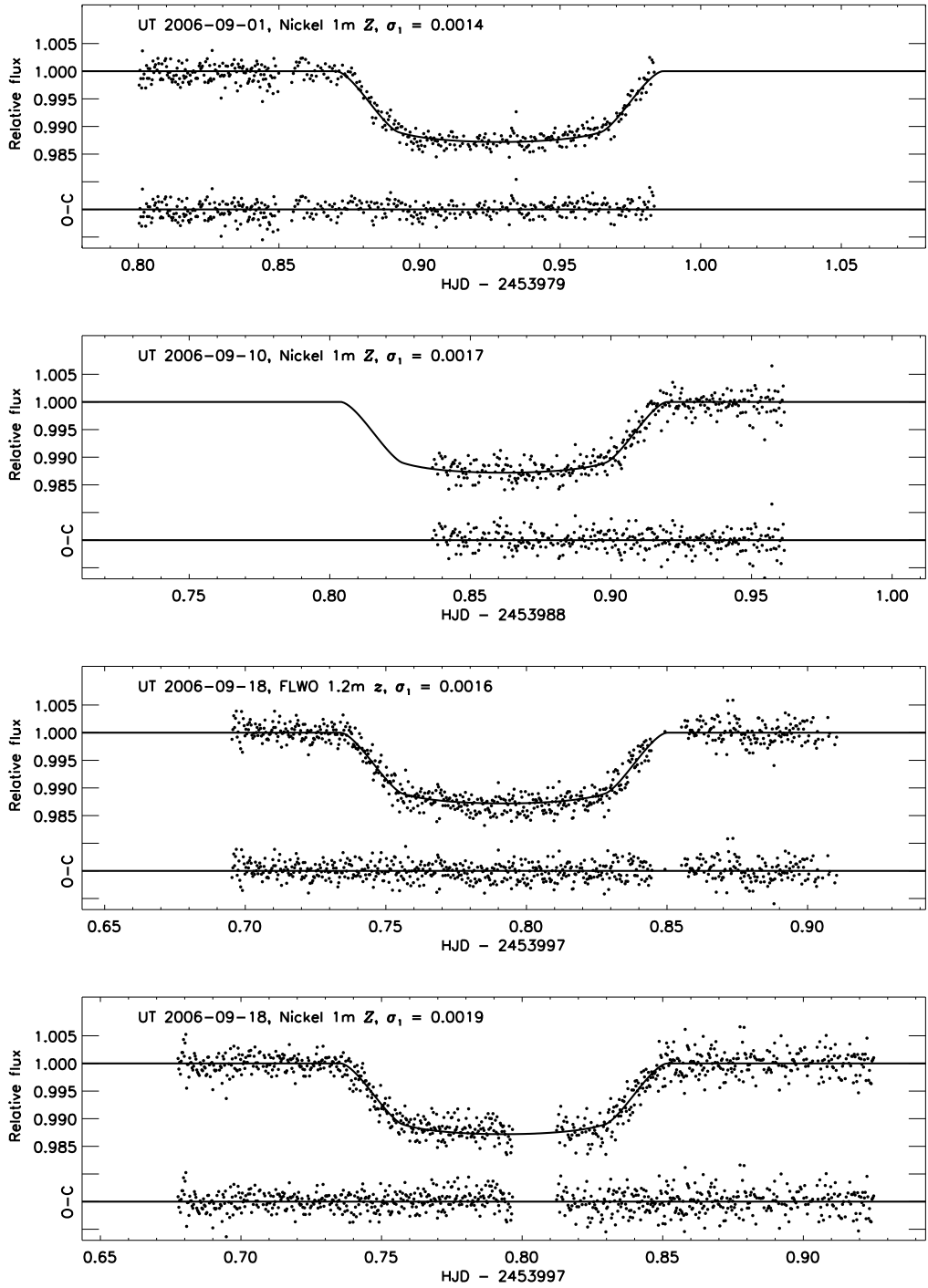}
\caption{ Relative photometry of HAT-P-1.
The residuals (observed$-$calculated) are plotted beneath the data.
\label{fig:1}}
\end{figure}

\begin{figure}[p]
\epsscale{0.85}
\plotone{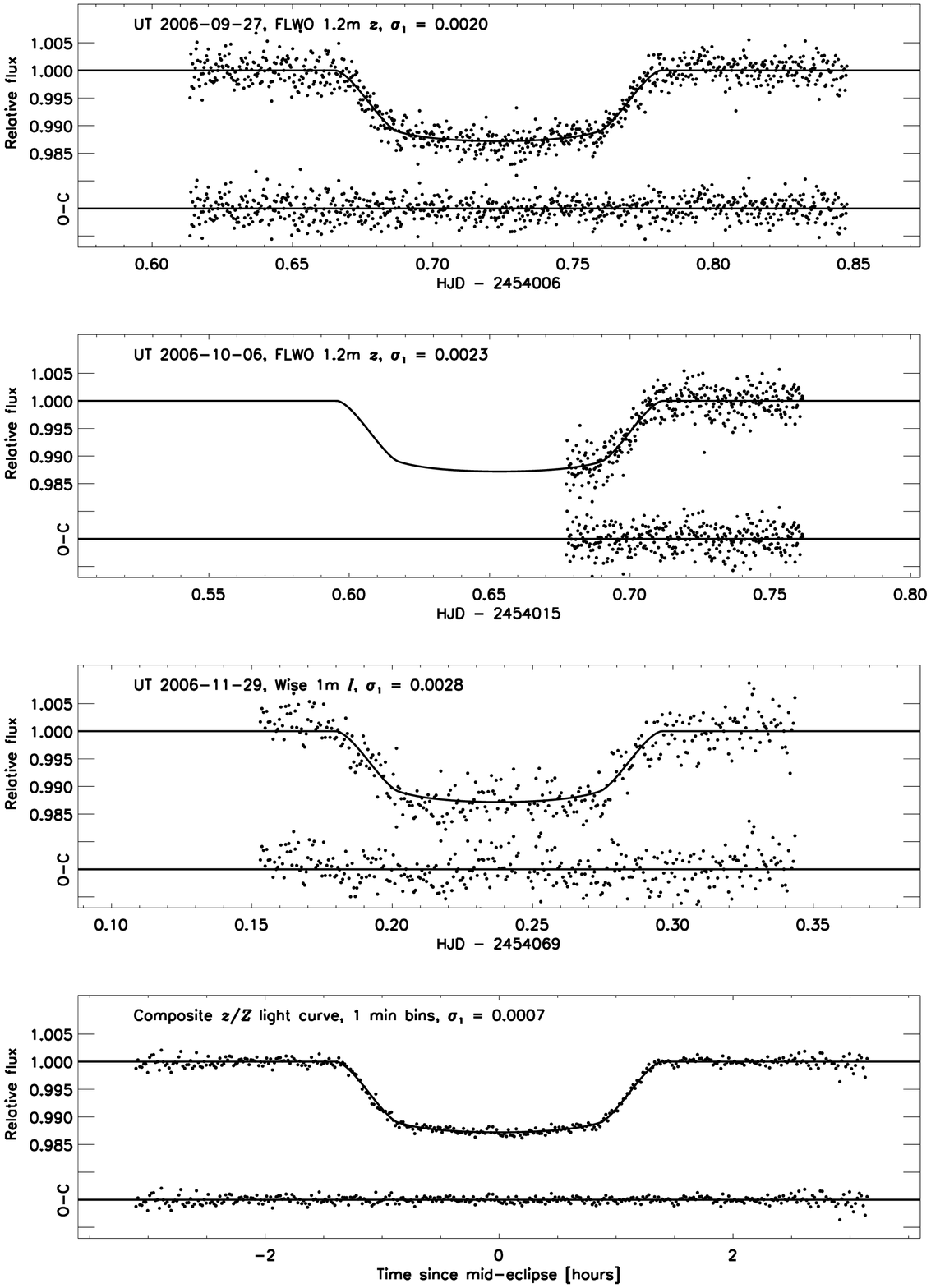}
\caption{ Relative photometry of HAT-P-1.
The residuals (observed$-$calculated) are plotted beneath the data.
The bottom panel is a composite light curve created from all of
the $z$ and $Z$ band data, by subtracting the best-fitting value of $T_c$ from
the time stamps of each light curve and then averaging into 1~min bins. 
\label{fig:2}}
\end{figure}

\section{Determination of System Parameters}

To estimate the planetary, stellar, and orbital parameters, and the
times of transit, we fitted a parameterized model to the photometric
data. The model and the fitting method were similar to those described
in previous TLC papers (see, e.g., Holman et al.~2006, Winn et
al.~2007b). It is based on a circular orbit\footnote{This is our
  default assumption in the absence of clear evidence for an eccentric
  orbit. Although the orbital fit of Bakos et al.~(2007) yielded the
  formal result $e=0.09\pm 0.02$, we regard this as only suggestive.
  The orbital eccentricity is subject to a positive bias in such fits,
  because $e$ is positive definite, and experience has shown that
  indications of a small nonzero eccentricity usually disappear after
  more velocity data are obtained.} of a star (with mass $M_\star$ and
radius $R_\star$) and a planet ($M_p$, $R_p$) about their center of
mass, inclined by an angle $i$ relative to the sky plane. Because one
of our goals was to measure the individual transit times, we allowed
each transit to have an independent value of $T_c$, the transit
midpoint, rather than forcing them to be separated by regular
intervals.

The most natural parameters one would like to know are $R_\star$ and
$R_p$, but these parameters cannot be determined independently from
the data. The relevant parameters that can be determined are
$R_p/R_\star$ and $R_\star/a$, where $a$ is the orbital semimajor
axis. The only property intrinsic to the star that follows directly
from the photometric data is the mean stellar density (see, e.g.,
Seager \& Mall\'{e}n-Ornelas~2003):
\begin{equation}
\rho_\star \equiv
\frac{M_\star}{\frac{4}{3}\pi R_\star^3} =
\frac{3\pi}{GP^2} \left( \frac{R_\star}{a} \right)^{-3}.
\end{equation}
In order to determine $R_\star$ and $R_p$ independently, one must have
an external estimate of $R_\star$, or $M_\star$, or some combination
of $R_\star$ and $M_\star$ besides $\rho_\star$. This external
estimate may come from supplementary observations such as the stellar
angular diameter and parallax (see, e.g., Baines et al.~2007), or from
the interpretation of the stellar spectrum with theoretical model
atmospheres and evolutionary tracks. The comparison with theory can be
facilitated by the estimate of $\rho_\star$ that comes from the
photometry, because the photometric estimate is often more precise
than the traditional spectroscopic gravity indicator, $\log g$ (see,
e.g., Pont et al.~2007, Sozzetti et al.~2007, Holman et al.~2007).

Our approach is to fix $M_\star$ at a fiducial value, and then
determine $R_\star$ and $R_p$ from the light curve. The scaling
relations $R_\star \propto M_\star^{1/3}$ and $R_p \propto
M_\star^{1/3}$ may then be used to estimate the resulting systematic
error due to the uncertainty in the stellar mass. This also makes it
easy to update the determinations of $R_\star$ and $R_p$ as our
understanding of the host star is revised through further observations
and analyses. In this case, we assumed $M_\star = 1.12$~$M_\odot$,
based on the analysis by Bakos et al.~(2007) in which the
spectroscopic properties of both members of the stellar binary were
fitted simultaneously to theoretical isochrones. The uncertainty in
$M_\star$ quoted by Bakos et al.~(2007) is 8\%, corresponding to a
systematic error of 2.7\% in our determinations of $R_\star$ and
$R_p$. The planetary mass $M_p$ hardly affects the photometric model
at all, since $M_p\ll M_\star$, but for completeness we used the
previously determined value $M_p=0.53$~$M_{\rm Jup}$.

To calculate the relative flux as a function of the projected
separation of the planet and the star, we employed the analytic
formulas of Mandel \& Agol~(2002) to compute the integral of the
intensity over the unobscured portion of the stellar disk. We assumed
the limb darkening law to be quadratic,
\begin{equation}
\frac{I_\mu}{I_1} = 1 - a(1-\mu) - b(1-\mu)^2,
\end{equation}
where $I$ is the intensity, and $\mu$ is the cosine of the angle
between the line of sight and the normal to the stellar surface. We
did not use the ``small-planet'' approximation. We fixed the
limb-darkening coefficients at the values calculated and tabulated by
Claret~(2004) for observations of a star with the observed spectral
properties.\footnote{Specifically, we used the tables for ATLAS
  models, interpolating for $T_{\rm eff} = 5975$~K, $\log g =
  4.45$~(cgs), log~[M/H]$=0.1$ and $v_t = 2.0$~km~s$^{-1}$. For the
  $z$ band, $a = 0.18$ and $b = 0.34$. We also used these values for
  the $Z$ band, finding it to provide a good fit. For the $I$ band,
  $a=0.22$ and $b=0.34$.} We also investigated the effects of changing
the limb-darkening law and fitting for the limb darkening parameters,
as discussed below. In addition, as mentioned in the previous section,
we fitted for the zero point and slope of the correlation between the
measured flux and the external variable that showed the strongest
correlation. For the FLWO data, this variable was airmass; for Lick it
was the column number; and for the Wise data it was the FWHM of
stellar images.

The fitting statistic was
\begin{equation}
\chi^2 =
\sum_{j=1}^{N_f}
\left[
\frac{f_j({\mathrm{obs}}) - f_j({\mathrm{calc}})}{\sigma_j}
\right]^2
,
\label{eq:chi2}
\end{equation}
where $f_j$(obs) is the flux observed at time $j$ and $\sigma_j$
controls the weights of the data points, and $f_j$(calc) is the
calculated flux. As noted in the previous section, the calculated flux
was the idealized flux of a transit light curve after subtracting a
linear function of a specified external variable.

For the data weights $\sigma_j$, many investigators use the calculated
Poisson noise, or the observed standard deviation of the
out-of-transit data. Experience has shown that these procedures
usually result in underestimated uncertainties in the model
parameters, because they neglect time-correlated errors (``red
noise''; see, e.g., Gillon et al.~2006), which are almost always
significant for ground-based data. In order to derive realistic
uncertainties on the parameters, it is important for $\sigma_j$ to
take red noise into account, at least approximately.

We did this as follows. The most relevant time scale is $\sim$20~min,
the ingress or egress duration. First we calculated $\sigma_1$, the
standard deviation of the unbinned out-of-transit data. (The results
for each light curve are given in Figs.~1 and 2.) Next we averaged the
out-of-transit data into 20~min bins, with each bin consisting of $N$
data points, depending on the cadence. Then we calculated the standard
deviation, $\sigma_N$. In the absence of red noise, we would observe
$\sigma_N = \sigma_1/\sqrt{N}$, but in practice $\sigma_N$ is larger
than $\sigma_1/\sqrt{N}$ by some factor $\beta$. Therefore, we set
the data weights equal to $\beta~\sigma_1$. The exact choice of
averaging time did not matter much. In the end, we used the mean value
of $\beta$ over averaging times ranging from 15 to 25
minutes. Typically we found $\beta\approx 2$, depending on the
telescope and sky conditions.\footnote{This procedure effectively
  increases the error bar of each measurement and results in a minimum
  value of $\chi^2/N_{\rm dof}$ that is smaller than unity. It is
  equivalent to setting $\sigma_j$ at the value that produce
  $\chi^2/N_{\rm dof}=1$ but then using $\Delta\chi^2=\beta^2$
  instead of $\Delta\chi^2=1$ to define the 68\% confidence limit.}

In all cases, to solve for the a posteriori probability distributions
of the model parameters, we used a Markov Chain Monte Carlo algorithm
[see, e.g., Tegmark et al.~(2004) for applications to cosmological
data, Ford (2005) for radial-velocity data, and Holman et al.~(2006)
or Burke et al.~(2007) for a similar approach to transit fitting].  We
ensured that the Gelman \& Rubin~(1992)~$R$ statistic was within 0.5\%
of unity, a sign of good mixing and convergence.  For each parameter,
we took the median value of the distribution to be our best estimate,
and the standard deviation as the 1~$\sigma$ uncertainty.

\section{Results}

The results are given in Tables~2 and 3. The first of these tables
gives the planetary, stellar, and orbital parameters, with the
fundamental parameters $R_\star/M_\star^{1/3}$, $R_p/M_p^{1/3}$, and
$i$ listed first. For the parameters that depend on the choice of
$M_\star$ (namely, $R_\star$, $R_p$, $a$, and $\rho_p$), we have
accounted for the systematic error due to the 8\% uncertainty in
$M_\star$. Table~3 gives the measured transit times.

\subsection{Planetary, Stellar, and Orbital Parameters}

We find the stellar radius to be $R_\star = 1.115 \pm 0.043~R_\odot$,
and the planetary radius to be $R_p = 1.203\pm 0.051~R_{\rm Jup}$. The
statistical error is comparable to the systematic error resulting from
the covariance with the stellar mass, implying that there is still
some scope for improvement through additional high-precision
photometry. Our value for the stellar radius agrees well with the
value $R_\star = 1.15^{+0.10}_{-0.07}~R_\odot$ determined by Bakos et
al.~(2007).  Those authors estimated $R_\star$ from an analysis of the
stellar spectrum---its effective temperature, surface gravity, and
metallicity---whereas we estimated $R_\star$ (actually
$R_\star/M_\star^{1/3}$) by fitting the transit light curves. The
agreement between these different methods of determining the stellar
radius is an important consistency check on both analyses. Our value
for the planetary radius is 1.5~$\sigma'$ smaller than the previously
determined value $R_p = 1.36^{+0.11}_{-0.09}~R_{\rm Jup}$, where
$\sigma'$ is the quadrature sum of the statistical errors of the two
estimates. Thus, we have revised the planetary radius downward and we
have improved the measurement precision by a factor of 2.

We performed a number of additional optimizations to check on the
sensitivity of the results to the choice of limb darkening function.
We tried replacing the quadratic law with either a linear law or the
four-parameter ``nonlinear'' law of Claret~(2000). For the quadratic
law, we tried replacing the ATLAS-based coefficients with the
PHOENIX-based coefficients of Claret~(2004). In none of these cases
did the optimized value of $R_p$ change by more than 0.5\% relative to
the value presented in Table~2. For the case of the linear law, we
tried fitting for the limb darkening coefficient rather than fixing it
at the prescribed value. In that case, $R_p$ increased by 1.8\%, which
is still small in comparison to the quoted error. (We found that the
present data are unable to constrain meaningfully more than one
limb-darkening parameter.) We conclude that the systematic error due
to the choice of limb darkening law is probably $\sim$1\%.

Also given in Table~2 are some results reproduced from Bakos et
al.~(2007) for convenience, as well as some useful derived
quantities. Among these quantities are the impact parameter $b$,
defined as $a\cos i/R_\star$ (where $a$ is the semimajor axis), the
radius ratio $R_p/R_\star$, the fractions $a/R_\star$ and $a/R_p$, and
the stellar mean density $\rho_\star$, which (as mentioned above) do
not depend on our choice for $M_\star$. We used the previous
measurement of the velocity semiamplitude of the spectroscopic orbit,
$K_\star=60.3\pm 2.1$~m~s$^{-1}$, to calculate the planetary surface
gravity, which is also independent of the undetermined stellar
properties (see, e.g., Southworth et al.~2007, Winn et al.~2007a).
The results for $a$ and the planetary mean density $\rho_p$ do depend
on the choice of stellar mass, and in those cases the quoted errors
have been enlarged appropriately to take this extra source of
uncertainty into account. For convenience in planning future
observations, we give the calculated values of the full transit
duration (the time between first and fourth contact, $t_{\rm IV} -
t_{\rm I}$), and the partial transit duration (the time between first
and second contact, or between third and fourth
contact).\footnote{Although the partial transit duration is listed as
  $t_{\rm II} - t_{\rm I}$ in Table~1, all of the results in Table~1
  are based on the entire light curves, including both ingress and
  egress data. Our model assumes $t_{\rm II} - t_{\rm I} = t_{\rm IV}
  - t_{\rm III}$.}

\subsection{Transit Times}

Table~3 gives the transit times measured from our data. We have used
these times to calculate a photometric ephemeris for this system.
Using only our 7 new measurements, we fitted a linear function of
transit epoch $E$,
\begin{equation}
T_c(E) = T_c(0) + E P.
\label{eq:ephemeris}
\end{equation}
The fit had $\chi^2/N_{\rm dof} = 1.6$ and $N_{\rm dof} = 5$,
suggesting that either the period is not exactly constant, or that the
transit time uncertainties have been underestimated.  Because one
would prefer to have an ephemeris with conservative error estimates
for planning future observations, we rescaled the measurement errors
by $\sqrt{1.6}$ and re-fitted the ephemeris, finding $T_c(0) =
2453997.79258(29)$~[HJD] and $P = 4.46543(14)$~days.  The numbers in
parentheses indicate the $1~\sigma$ uncertainty in the final two
digits. Our derived period agrees with the value $4.465290(90)$~days
determined by Bakos et al.~(2007), within the respective 1~$\sigma$
limits. Figure~3 is the O$-$C (observed minus calculated) diagram for
the transit times.

\begin{figure}[p]
\epsscale{1.0}
\plotone{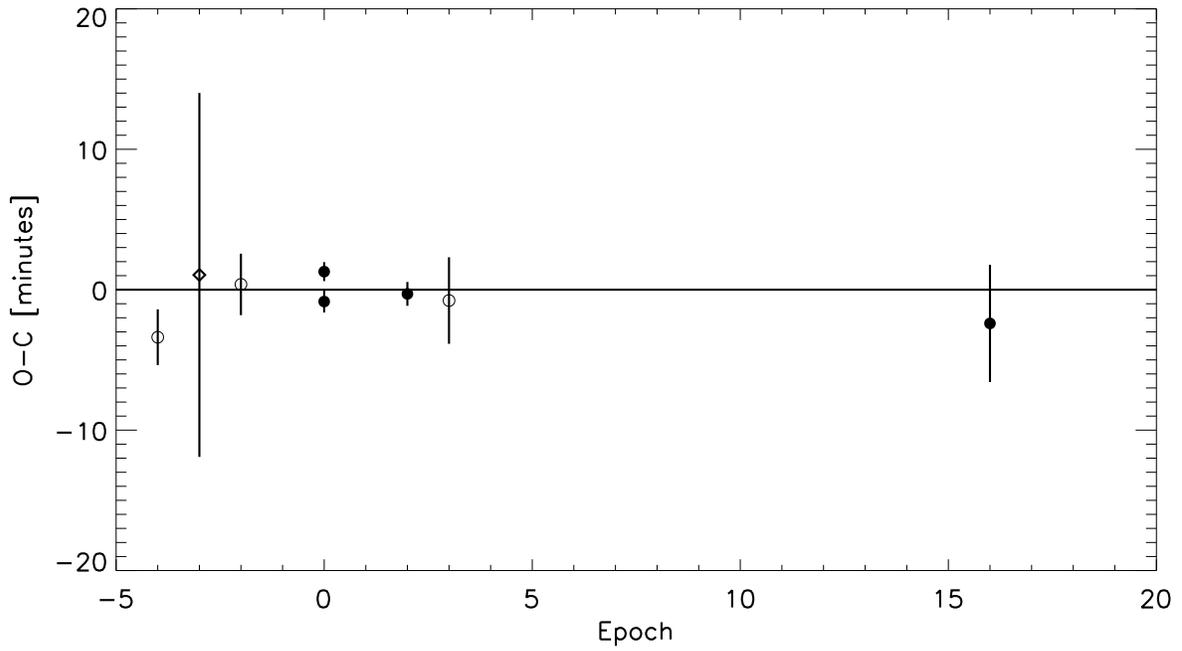}
\caption{ Transit timing residuals for HAT-P-1b. The calculated times,
  using the ephemeris derived in \S~4.2, have been subtracted from the
  observed times. The filled symbols represent observations of
  complete transits. Open symbols represent observations of partial
  transits. The diamond represents the previous observation by Bakos
  et al.~(2007), which was not used in the fit but which agrees well
  with the computed ephemeris.
  \label{fig:3}}
\end{figure}

\section{Summary and Discussion}

We have presented new photometry of HAT-P-1 spanning the times of
transit of its close-in giant planet. The photometry improves the
precision with which the stellar and planetary radii are known by a
factor of 2, and places the measurements on a more robust footing by
determining the stellar mean density directly from the photometric
data. We have also updated the transit ephemeris, to help in planning
future observations.

Although the revised planetary radius is somewhat lower than the
previously determined value, the planet HAT-P-1b is still among the
largest and least dense of the known transiting exoplanets. Its mean
density ($0.376\pm 0.031$~g~cm$^{-3}$) is comparable to that of the
famously oversized planet HD~209458b ($0.35$~g~cm$^{-3}$; Knutson et
al.~2007b). A third planet with a comparably low mean density is
WASP-1b ($0.36$~g~cm$^{-3}$; Collier Cameron et al.~2007; Charbonneau
et al.~2007; Shporer et al.~2007).

There is an extensive literature on the interpretation of exoplanetary
radii, and in particular on the subject of these apparently
``bloated'' planets. This term refers to the apparent conflict (of
order 10-20\%) between the measured planetary radii and the calculated
radii using simple structural models for hydrogen/helium planets of
the appropriate mass, temperature, age, and degree of external heating
by the parent star. Many mechanisms have been proposed to sustain
hotter gaseous envelopes and therefore larger planets: the efficient
delivery of heat from the star to the planetary interior (Guillot \&
Showman~2002; Showman \& Guillot~2002); the production of internal
heat by tidal interactions (Bodenheimer et al.~2003; Winn \&
Holman~2005); and the trapping of internal heat by enhanced
atmospheric opacities (Burrows et al.~2007) or inhibited convection
(Chabrier \& Baraffe~2007). Only the tidal mechanisms have been
specific or predictive enough to be ruled out; the obliquity-tide
theory of Winn \& Holman~(2005) did not withstand more detailed
dynamical analysis (Levrard et al.~2007, Fabrycky et al.~2007), and
the eccentricity-tide mechanism of Bodenheimer et al.~(2003) does not
seem to be operative in the case for which it was invented, HD~209458b
(Laughlin et al.~2005). Which (if any) of the other mechanisms
contribute to the observed radii of transiting exoplanets is not yet
clear.

However, for HAT-P-1b, this issue may be a red herring. We find that
there is no strong conflict with structural models at this point, as
long as the planet does not have a massive core of heavy
elements. Burrows et al.~(2007) recently computed models for many of
the transiting exoplanets including HAT-P-1b in particular, taking
into account the appropriate planetary mass, orbital distance, stellar
luminosity, stellar spectrum, and stellar age. Assuming a planet of
solar composition, an atmosphere of standard solar-composition
opacity, and no dense heavy-element core, they calculated $R_p =
1.18$--$1.22~R_{\rm Jup}$ over the plausible age range $3.5\pm
1.0$~Gyr. This range of calculated values for $R_p$ comfortably
overlaps the 1~$\sigma$ range in our measured value, $1.203\pm
0.051$~$R_{\rm Jup}$.

Fortney, Marley, \& Barnes~(2007) have also provided theoretical
estimates of exoplanetary radii over a wide range of conditions,
although they are not specifically tuned for any particular cases of
the known exoplanets. For a coreless H/He planet with mass $0.5~M_{\rm
  Jup}$ orbiting a 3-Gyr-old solar-luminosity star at a distance of
0.045~AU (and thereby receiving nearly the same flux as HAT-P-1, which
orbits a $\sim$1.5~$L_\odot$ star at a distance of 0.055~AU), Fortney
et al.~(2007) predict a planetary radius $R_p=1.12$~$R_{\rm
  Jup}$. This is smaller than the value computed by Burrows et
al.~(2007), and at least part of the reason for the difference is that
Fortney et al.~(2007) did not account for the ``transit radius
effect'': the enlarged size of the transit-measured radius relative to
the $\tau=2/3$ photosphere that is usually taken to be the ``radius''
by theoreticians. This effect amounts to a few per cent in the
planetary radius (see also Burrows et al.~2003). Assuming that this
effect adds between 0\% and 5\% to the calculated radius, the
difference between the calculated radius of $R_p$ and our measured
value is 0.5--1.6~$\sigma$, i.e., not very significant.

We conclude that the present data are consistent with current models
of coreless, solar-composition, strongly irradiated giant planets.
Bakos et al.~(2007) estimated a stellar metallicity of $Z=0.025$,
i.e., comparable to the Sun, and hence the inference of a small or
absent core is broadly consistent with the core-metallicity relation
proposed by Guillot et al.~(2006). Of course, there are many other
possibilities that are also consistent with the data, such as a planet
with a dense core that also has either an extra source of atmospheric
opacity or an extra source of internal heat. As of now there is no way
to distinguish among these possibilities.

As discussed by Burrows et al.~(2007), it is becoming clear that there
are many determinants of planetary radii. By considering the entire
ensemble of exoplanets one can fully appreciate the strengths and
weaknesses of theoretical models, and possibly obtain clues about
interesting processes that may have been overlooked. This requires not
only the discovery of new transiting systems, but also high-precision
determinations of the system parameters, such as the present study.

\acknowledgments We thank Debra Fischer and Geoff Marcy for helpful
discussions, and John Southworth for his publicly available code for
finding limb darkening parameters. A.P.\ is grateful for the
hospitality of the Harvard-Smithsonian Center for Astrophysics, where
some of this work was carried out. M.J.H.\ acknowledges support for
this work from NASA Origins grant NG06GH69G. G.\'{A}.B.\ was supported
by NASA through a Hubble Fellowship Grant HST-HF-01170.01. P.K.G.W.\
was supported by an NSF Graduate Student Research Fellowship.
KeplerCam was developed with partial support from the Kepler Mission
under NASA Cooperative Agreement NCC2-1390 (PI~D.~Latham), and the
Keplercam observations described in this paper were partly supported
by grants from the Kepler Mission to SAO and PSI.

\begin{deluxetable}{lcccc}
\tabletypesize{\normalsize}
\tablecaption{Photometry of HAT-P-1\label{tbl:photometry}}
\tablewidth{0pt}

\tablehead{
\colhead{Telescope} &
\colhead{Filter} &
\colhead{Heliocentric Julian Date} & 
\colhead{Relative flux}
}

\startdata
      FLWO & $  z$ & $  2453997.69528$ & $         0.9988$ \\
      FLWO & $  z$ & $  2453997.69560$ & $         1.0006$ \\
      FLWO & $  z$ & $  2453997.69591$ & $         1.0031$ \\
      FLWO & $  z$ & $  2453997.69621$ & $         1.0009$ \\
      Lick & $  Z$ & $  2453997.91750$ & $         1.0013$ \\
      Lick & $  Z$ & $  2453997.91781$ & $         0.9965$ \\
      Lick & $  Z$ & $  2453997.91811$ & $         1.0016$ \\
      Wise & $  I$ & $  2454069.33838$ & $         1.0014$ \\
      Wise & $  I$ & $  2454069.33894$ & $         0.9970$ \\
      Wise & $  I$ & $  2454069.33950$ & $         1.0000$
\enddata 

\tablecomments{The time stamps represent the Heliocentric Julian Date
  at the time of mid-exposure. We intend for this Table to appear in
  entirety in the electronic version of the journal. Excerpts are
  shown here to illustrate its format. The data are also available
  from the authors upon request.}

\end{deluxetable}

\begin{deluxetable}{lcc}
\tabletypesize{\normalsize}
\tablecaption{System Parameters of HAT-P-1b\label{tbl:params}}
\tablewidth{0pt}

\tablehead{
\colhead{Parameter} & \colhead{Value} & \colhead{Uncertainty}
}

\startdata
            $(R_\star/R_\odot)/(M_\star/1.12~M_\odot)^{1/3}$& $          1.115$ & $          0.034$ \\
            $(R_p/R_{\rm Jup})/(M_\star/1.12~M_\odot)^{1/3}$& $          1.203$ & $          0.043$ \\
                                                   $i$~[deg]& $          86.22$ & $           0.24$ \\
\hline
                                       $M_\star$~[$M_\odot$]& $           1.12$ & $           0.09$ \\
                                       $M_p$~[$M_{\rm Jup}$]& $           0.53$ & $           0.04$ \\
              Velocity semiamplitude, $K_\star$ [m~s$^{-1}$]& $           60.3$ & $            2.1$ \\
                                  Orbital period, $P$~[days]& $        4.46529$ & $        0.00009$ \\
\hline
                                       $R_\star$~[$R_\odot$]& $          1.115$ & $          0.043$ \\
                                       $R_p$~[$R_{\rm Jup}$]& $          1.203$ & $          0.051$ \\
                                             $R_p / R_\star$& $        0.11094$ & $        0.00082$ \\
                                    Semimajor axis, $a$~[AU]& $         0.0551$ & $         0.0015$ \\
                                  $b \equiv a\cos i/R_\star$& $          0.701$ & $          0.023$ \\
                                                 $a/R_\star$& $          10.64$ & $           0.32$ \\
                                                     $a/R_p$& $           95.9$ & $            3.5$ \\
                               $t_{\rm IV} - t_{\rm I}$~[hr]& $          2.779$ & $          0.032$ \\
                               $t_{\rm II} - t_{\rm I}$~[hr]& $          0.508$ & $          0.035$ \\
                                  $\rho_\star$~[g~cm$^{-3}$]& $           1.14$ & $           0.10$ \\
                                      $\rho_p$~[g~cm$^{-3}$]& $          0.376$ & $          0.031$ \\
                                  $GM_p/R_p^2$ [cm~s$^{-2}$]& $          904.5$ & $           66.1$
\enddata

\tablecomments{This table has three sections, divided by horizontal
  lines.  The top section lists the parameters that were estimated by
  fitting the new photometric data, as explained in \S~3.  The orbital
  eccentricity was assumed to be exactly zero.  The middle section
  lists some parameters from Bakos et al.~(2007), reproduced here for
  convenience. The bottom section lists some interesting parameters
  that can be derived from the parameters in the first two sections.}

\end{deluxetable}

\begin{deluxetable}{lcccc}
\tabletypesize{\normalsize}
\tablecaption{Mid-transit times of HAT-P-1\label{tbl:times}}
\tablewidth{0pt}

\tablehead{
\colhead{Observatory} & \colhead{Epoch} & \colhead{Mid-transit time} & \colhead{Uncertainty} \\
                      & \colhead{$E$}   & \colhead{[HJD]}            & \colhead{[days]}      
}

\startdata
      Lick & $  -4$ & $ 2453979.92848$ & $0.00069$ \\
      Lick & $  -2$ & $ 2453988.86197$ & $0.00076$ \\
      FLWO & $   0$ & $ 2453997.79200$ & $0.00054$ \\
      Lick & $   0$ & $ 2453997.79348$ & $0.00047$ \\
      FLWO & $   2$ & $ 2454006.72326$ & $0.00059$ \\
      FLWO & $   3$ & $ 2454011.18837$ & $0.00107$ \\
      Wise & $  16$ & $ 2454069.23795$ & $0.00290$ \\
\enddata

\tablecomments{Based on these new measurements, we derived a transit
  ephemeris $T_c(E) = T_c(0) + EP$ with $T_c(0) =
  2453997.79258(29)$~[HJD] and $P = 4.46543(14)$~days, where the
  numbers in parentheses indicate the 1$\sigma$ uncertainty in the
  final two digits. We note that Bakos et al.~(2007) derived a more
  precise period based on observations over 217 cycles,
  $P=4.465290(90)$~days.}

\end{deluxetable}

\end{document}